\documentclass[twocolumn,amssymb,prx,superscriptaddress,longbibliography]{revtex4-2}
\usepackage[colorlinks=true,linkcolor = red,citecolor=blue,urlcolor=blue,anchorcolor = blue]{hyperref}
\usepackage{array}
\usepackage{lipsum}
\usepackage{amsmath}
\usepackage{amsbsy}
\providecommand{\selectlanguage}[1]{}
\usepackage{amssymb}
\usepackage{amsthm}
\usepackage{bigstrut}
\usepackage{ulem}
\usepackage{graphicx} 
\usepackage{tabularx}
\usepackage{tikz}
\usepackage{braket}
\usepackage{multirow}
\usepackage[figuresright]{rotating}
\usepackage{float}
\usepackage{ulem}
\usepackage{esint} 
\usepackage{mathrsfs} 
\usepackage{physics}
\usepackage{mathtools} 
\usepackage[perpage]{footmisc}
\usepackage{tensor} 
\usepackage{siunitx} 
\newcommand{\RNum}[1]{\uppercase\expandafter{\romannumeral #1\relax}}

\newcommand{\e}{\text e}
\newcommand{\ee}{\text{\it{e}}} 
\newcommand{\ii}{\text i}

\newcommand{\level}[1]{^{(#1)}}

\newcommand{\down}{\downarrow}

\renewcommand{\Re}{\operatorname{Re}}
\renewcommand{\Im}{\operatorname{Im}}
\newcommand{\beq}{\begin{eqnarray} }
\newcommand{\eeq}{\end{eqnarray} }
\newcommand{\Beq}{\begin{eqnarray*} }
\newcommand{\Eeq}{\end{eqnarray*} }
\newcommand{\Bmat}{\left(\begin{matrix}}
\newcommand{\Emat}{\end{matrix}\right)}

\newcommand{\bit}{\begin{itemize} }
\newcommand{\eit}{\end{itemize} }
\newcommand{\ben}{\begin{enumerate} }
\newcommand{\een}{\end{enumerate} }

\begin{document}

\title{Super-Solid phase in a $U(2)$ symmetric $S=1$ Magnet on the Triangular Lattice}

\author{Si-Cheng Wang}
\affiliation{Department of Physics and Beijing Key Laboratory of Opto-electronic Functional Materials and Micro-nano Devices, Renmin University of China, Beijing, 100872, China}
\affiliation{Beijing National Laboratory for Condensed Matter Physics, Institute of Physics, Chinese Academy of Sciences, Beijing 100190, China}
\affiliation{University of Chinese Academy of Sciences, Beijing 100049, China}
\author{Zheng-Xin Liu }
\email{liuzxphys@ruc.edu.cn}
\affiliation{Department of Physics and Beijing Key Laboratory of Opto-electronic Functional Materials and Micro-nano Devices, Renmin University of China, Beijing, 100872, China}
\affiliation{Key Laboratory of Quantum State Construction and Manipulation (Ministry of Education), Renmin University of China, Beijing, 100872, China}

\date{\today}

\begin{abstract} A spin supersolid is characterized by the simultaneous breaking of lattice translation and continuous spin rotation symmetries. In this work, we study a spin-1 model with $U(2)\cong SU(2)\times U(1)/Z_2$ symmetry on the triangular lattice, and determine the phase diagram with a variational $\mathbb CP^2$ approach. We identify a novel supersolid phase which contains a 3-sublattice solid order and a superfluid order with spontaneous $SU(2)$-symmetry breaking. Unlike usual supersolid phases having 
only one Goldstone mode, the $SU(2)$-supersolid phase has 
two Goldstone modes. Another important feature of this supersolid is that the 
spin excitation spectrum has symmetry protected double degeneracy in the whole Brillouin zone. 
As by-products, several other ordered phases are obtained, including the ferromagnetic and the antiferromagnetic states breaking the $SU(2)$ symmetry, as well as genuine phases that completely break the $U(2)$ symmetry. Furthermore, the instabilities of $SU(3)$-flavor linear spin-wave theory are consistent with the phase boundaries between different ordered phases. \\[-8pt]


\end{abstract}

\maketitle

\section{INTRODUCTION}

The concept of supersolidity has attracted considerable interest in condensed matter physics since its proposal in 1970\cite{Leggett_Supersolid_1970}. A supersolid is an exotic phase of matter characterized by the simultaneous coexistence of Bose-Einstein condensation (BEC) and density wave order, spontaneously breaking both the $U(1)$ particle number conservation symmetry and continuous translational symmetry. However, the existence of such a phase in continuum Bose gases remained controversial for many years\cite{Kim_SupersolidHelium_2004,Kim_SupersolidHeliumAbsence_2012,Boninsegni_RevSuperSolid_2012}.

Later, it was proposed that the lattice version of the supersolid phase can indeed arise for hard-core bosons (bosons having infinite on-site repulsion) in the extended Hubbard model on frustrated lattices with nearest-neighbor interactions\cite{Murthy_MFSupersolid_1997}. Since the hard core Boson model can be exactly mapped into a spin-1/2 XXZ system, the resultant supersolid phase is essentially a spin super-solid, namely, a magnetic phase with the emergence of spin-superfluid order (i.e. in-plane magnetic order) that simultaneously breaks the original lattice translation symmetry (characterized by enlarged unit cells). This scenario for realizing super-solids in hard-core bosons or spin systems was first illustrated via mean field theory\cite{Murthy_MFSupersolid_1997} and later verified by a series of numerical studies \cite{Heidarian_PersistentSupersolidHardCoreBoson_2005, Melko_SuperSolidOrderDisorder_2005, Wessel_SuperSolidHardCoreBoson_2005, Boninsegni_SuperSolidHCBosonTriangle_2005, Fa_AFMSupersolid_2009,Jiang_Supersolid_2009,Wang_ExtendedSupersolid_2009,Heidarian_SupersolidVariational_2010,Yamamoto_XXZ_2014,Yamamoto_LayeredTLAF_2015,Sellmann_XXZ_2015,Yamamoto_XXZ_2019}. The spin-supersolid order enriches the family of magnetic phases of matter and has attracted the interest of the experimental studies. Recently, progress have been made in the realization of the spin super-solid phases in the material Na$_2$BaCo(PO$_4)_2$ \cite{GaoYuan_SpinSuperSolidNBP_2022, Xiang_GiantMagnetocloric_2024}.

On the other hand, magnets carrying spin larger than $1/2$ can also support exotic phases of matter. For instance, many quantum phases exist in spin-1 systems, such as the Haldane chain\cite{HaldanePLA1983, HaldanePRL1983, Buyers-1986} and general symmetry protected topological phases\cite{GuWen2009, Pollmann2010, ChenGuLiuWen2011, ChenGuWen2011_1D, LiuChenWen2011, LiuZhouTuNgWen2012, LiuMeiYeWen14}, chiral spin liquids\cite{Thomale09, LiuTuWuZhouNg18, ShiZhangLiu23, Xiang25}, and so on. Like the mapping between spin-1/2 systems and hard-core bosons, a spin-1 magnet can also be considered as a bosonic \textcolor{blue}{system} in which the hard-core constraint is relaxed to $0\le \expval{n_i}\le 2$ \cite{Batista_GeneralizedJordanWigner_2001, Batista_AlgebraicInteractingQuantum_2004, Sengupta_SupersolidSpinOne_2007}. Since the 
3-dimensional bosonic Hilbert space supports eight linearly independent Hermitian operators (other than the identity matrix), generally one should consider the three states of spin-1 as the basis carrying the fundamental representation of the $SU(3)$ group. Consequently, many classical orderes can exist for spin-1 systems, including the magnetic dipole order, quadrupole order (also called spin nematic order)\cite{Tsunetsugu_NematicOrder_2006,Senthil06, LiuZhouNg2011_FermionMF, LiuZhouNg2011_Spin-1SL, Zhou19}, and even  $U(1)$ spin supersolid phases \cite{Sengupta_SpinSupersolidChain_2007, Sengupta_SupersolidSpinOne_2007,  Sheng_QuadrupoleWaveSpinNematics_2025}.

In this work, we study a triangular lattice spin-1 model with $U(2)$ symmetry.  Using a semiclassical approach, we obtain a supersolid phase which spontaneously breaks the $SU(2)\subset U(2)$ symmetry as well as the lattice translation symmetry. This supersolid phase contains collinear `antiferromagnetic' (AFM) order in the $SU(2)$ sector and two Goldstone modes. Besides, we found several other ordered phases, including a solid phase, a $SU(2)$-`ferromagnetic' (FM) phase, a $SU(2)$-120$^\circ$ `antiferromagnetic'(AFM) phase, and other spin nematic phases which completely break the $U(2)$ symmetry. Furthermore the $SU(3)$ spin wave spectrum indicates that the instability of the `magnons' agree with the phase boundary of the semiclassical phase diagram.

The rest part of the paper is organized as follows. In section \ref{sec2:Model}, we introduce the model and its symmetries as well as semi-calssical variational wave functions. The phase diagram, phase transitions and the properties of each phase are given in \ref{sec3:phasediag}. In section \ref{sec4:Su3spinwave}, the spin wave spectrum are provided, especially the magnon bands of the $SU(2)$ supersolid phase are shown to have symmetry protected double degeneracy. Section \ref{sec5:sum} is devoted to summary and discussions. 

\section{Model and Variational States}\label{sec2:Model}

\subsection{Spin-1 model with $U(2)$ symmetry}

We start from the following SU(3) Heisenberg model which has the highest symmetry for spin-1 systems,
\begin{align}\label{su3}
    H_{\rm su3}=J_3\sum_{\expval{ij}}\sum_{\alpha=1}^8\lambda^\alpha_{i}\lambda^\alpha_{j},
\end{align}
where $\lambda^1,..., \lambda^8$ are the eight Gell-Mann matrices which generate the SU(3) group. Here we slightly modify the form of the Gell-Mann matrices $\lambda^3,\lambda^8$ as 
\Beq
\lambda^3=\Bmat1&0&0\\0&0&0\\0&0&-1\Emat, \lambda^8={1\over\sqrt3}\Bmat1&0&0\\0&-2&0\\0&0&1\Emat,
\Eeq
and keep the rest ones unchanged:
\Beq
\lambda^1=\Bmat0&1&0\\1&0&0\\0&0&0\Emat, \lambda^2=\Bmat0&-i&0\\i&0&0\\0&0&0\Emat,\\
\lambda^4=\Bmat0&0&1\\0&0&0\\1&0&0\Emat, \lambda^5=\Bmat0&0&-i\\0&0&0\\i&0&0\Emat,\\
\lambda^6=\Bmat0&0&0\\0&0&1\\0&1&0\Emat, \lambda^7=\Bmat0&0&0\\0&0&-i\\0&i&0\Emat.\\
\Eeq

Notice that the nonzero entries of the generators $ \lambda_4, \lambda_5,\lambda_3,$ are essentially three Pauli matrices $\sigma^x,\sigma^y,\sigma^z$ within the subspace spanned by the first two bases $\ket{1},\ket{-1}$ of the spin-1 Hilbert space, and generate an $SU(2)$ subgroup of $SU(3)$. Furthermore, since $\lambda^8$ commutes with the three generators $\lambda_3,\lambda_4,\lambda_5$, these four generators together generate the complete symmetry group $U(2)=SU(2)\times U(1)/Z_2$ of the 2-dimensional subspace spanned by $\ket{1},\ket{-1}$, where the subgroup $U(1)$ is generated by $\lambda^8$ and the spin parity group $Z_2=\{E, P_s\}$ is the common center of $SU(2)$ and $U(1)$ with the {\it spin parity} defined by 
$$P_s=e^{-i\lambda^3\pi}=e^{-i\lambda^4\pi}=e^{-i\lambda^5\pi}=(-1)^{S_z}.
$$
Actually, the four operators of $U(2)$ can be expressed in forms of $S=1$ spin operators as 
\Beq
&&\lambda^4=S_x^2-S_y^2, \ \lambda^5=S_{xy}=S_xS_y+S_yS_x,\\
&&\lambda^3=S_z,\  \lambda^8={1\over\sqrt3}(3S_z^2-2).
\Eeq
Similarly, the rest generators can also be written in forms of spin operators as $\lambda^{1,6}={1\over\sqrt2}(S_x\pm S_{xz}), \lambda^{2,7}={1\over\sqrt2}(S_y\pm S_{yz})$.

The above $U(2)$ symmetry allows additional interactions, including the $J_2$ term in the $SU(2)$ sector,
\begin{align}\label{u2_1}
    H_{\rm su2}=J_2\sum_{\expval{ij}}\sum_{m=3}^5\lambda_{i}^m\lambda_{j}^m,
\end{align}
and the $\lambda_8$-term in the $U(1)$ sector,
\begin{align}\label{u2_2}
    H_{\rm u1}=J_1\sum_{\expval{ij}}\lambda^8_{i}\lambda^8_{j}.
\end{align}

The model we will study in the present work is a sum of the three terms in (\ref{su3}), (\ref{u2_1}), (\ref{u2_2}):
\beq\label{Ham}
    H=H_{\rm u1}+H_{\rm su2}+H_{\rm su3}.
\eeq
We will only consider the `anti-ferromagnetic' coupling with $J_1>0$, and will fix $J_3=1$ in later discussion. 

Besides the $U(2)\cong SU(2)\times U(1)/Z_2 \subset SU(3)$ symmetry \cite{Zhou19}, the above model also preserves lattice translation symmetry, spatial inversion symmetry $P$, time reversal symmetry 
\beq\label{symT}
T=e^{-i{\lambda_5\over2}\pi}K =e^{-i{S_{xy}\over2}\pi}K =\Bmat&&-1\\&1&\\1&&\Emat K,
\eeq 
and $D_6$ point group symmetry, with 
$$
P^2=1, \ T^2=P_s={\rm diag}(-1,1,-1),
\ T^4=1
$$ 
and 
\Beq
&&T\lambda^{8}T^{-1}=\lambda^{8},\ \ T\lambda^{3,4,5}T^{-1}=-\lambda^{3,4,5},\\ 
&&T\lambda^{1,2}T^{-1}=\lambda^{6,7},\ \ T\lambda^{6,7}T^{-1}=-\lambda^{1,2}.
\Eeq

\subsection{Semiclassical Variational Wave functions}\label{sec:H}
Any state in the Hilbert space of a single spin-1 can be written as
\begin{align}
    \ket{\boldsymbol{d}}=d_+\ket{1}+d_0\ket{0}+d_-\ket{-1}
\end{align}
where $\ket{1},\ket{0},\ket{-1}$ are the eigenstates of $\hat S^z$ with eigenvalue $1,0,-1$, respectively. The coefficients $d_+,d_0,d_-$ are complex numbers subject to the normalization constraint $ |d_+|^2+|d_0|^2+|d_-|^2=1$.

Since an overall phase shift does not yield a distinct physical state, the complete manifold of a spin-1 state is 
$SU(3)/U(2)\cong \mathbb CP^2$. A feasible parameterization scheme of this manifold is given by
\Beq
	&& d_+=\theta_2^{1/4}\theta_1^{1/2}e^{i\phi_1},\\
	&& d_0=\theta_2^{1/4}\sqrt{1-\theta_1}e^{i\phi_2},\\
       && d_-=\sqrt{1-\theta_2^{1/2}},        
\Eeq
where $0\le \theta_1,\theta_2\le 1$ and $0\le \phi_1,\phi_2\le 2\pi$ are $\mathbb CP^2$ parameters chosen from a uniform distribution\cite{George_Sphere_1972}. Within this classical framework, we adopt a variational wave function for the spin-1 many-body state in a direct product form
\beq\label{TrialState}
    \ket{\Psi}=\bigotimes_{i=1}^N\ket{\boldsymbol{d}_i(\theta_{1i},\theta_{2i},\phi_{1i},\phi_{2i}}).
\eeq
This is so-called $\mathbb CP^2$ framework, as employed in Ref.\cite{Remund_CP2MC_2022}. Within this variational wavefunction, the expectation values of the order parameters on each site are
$\expval{\lambda_{4,i}}=2\Re\qty(d_{+i}^*d_{-i}), \expval{\lambda_{5,i}}=2\Im\qty(d_{+i}^*d_{-i}), \expval{\lambda_{3,i}}=|d_{+i}|^2-|d_{-i}|^2, \expval{\lambda_{8,i}}={1\over\sqrt3}\qty(1-3|d_{0i}|^2).$
The energy of the trial state (\ref{TrialState}) is then given by 
\begin{align}
    \nonumber
E=&2J_3\sum_{\expval{ij}}\qty(|{d}_{+i}^*{d}_{+j} + d_{0i}^*{d}_{0j} + d_{-i}^*{d}_{-j}|^2-\dfrac{2}{3})
    \\
    \label{Total energy functional for U2 model}
    &+J_2\sum_{\expval{ij}}\sum_{m=3}^5\expval{\lambda_i^m}\expval{\lambda_j^m}+J_1\sum_{\expval{ij}}\expval{\lambda_i^8}\expval{\lambda_j^8}.
\end{align}

\begin{figure}[t]
    \centering
    \includegraphics[width=1\linewidth]{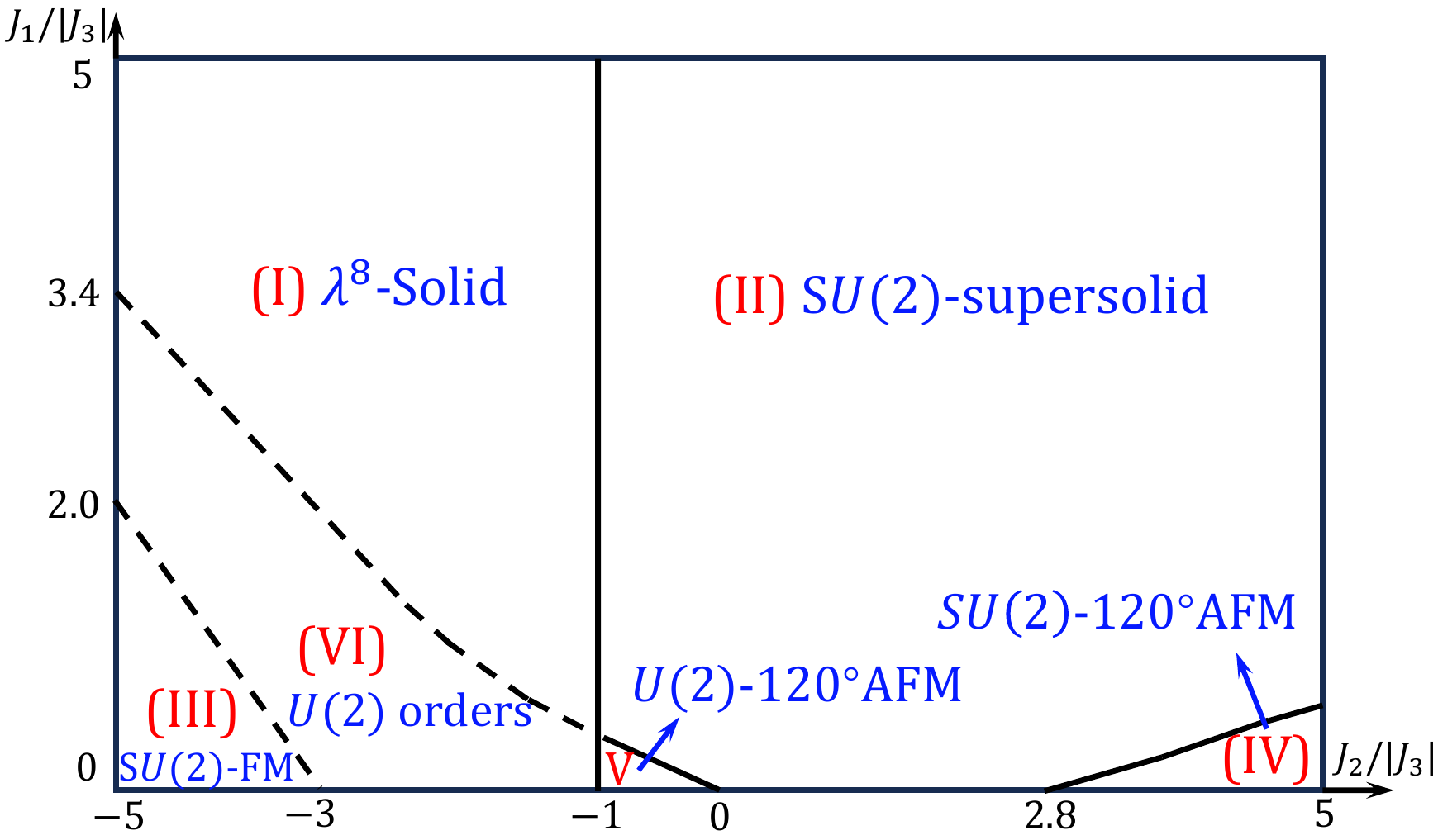}
    \caption{The phase diagram of the model (\ref{Ham}) with $J_1>0$. There are totally six different phases : the $\lambda^8$-solid, the $SU(2)$-supersolid, the $SU(2)$-FM, the $SU(2)$-$120^{\circ}$AFM, the $U(2)$-$120^{\circ}$AFM, and the $U(2)$ orders. The solid line signifies the first-order transition, and the dashed line labels the second-order transition.}
    \label{fig:0}
\end{figure}

The trial state (\ref{TrialState}) generally contains long-range order. Assuming that the unit cell contains $n_\Lambda$ sites\cite{IntroFrustratedMagnetism, Lauchli_Quadrupolar_2006}, then the variational wave function can also be written as 
\beq\label{PsiN}
    \ket{\Psi}=\prod_{j=1}^{n_{\Lambda}}\prod_{i\in\Lambda_j}\ket{\boldsymbol{d}_j}_i
\eeq
where $i$ is a site index, $\Lambda_j$ denotes the set of sites belonging to the $j$-th sublattice, and the on-site wave function is assumed to be identical for all spins within the same sublattice(i.e. $\ket{\boldsymbol{d}_j}_{i_1}=\ket{\boldsymbol{d}_j}_{i_2}=\cdots$). Thus, the task of finding ground state reduces to optimizing the enery functional \eqref{Total energy functional for U2 model} with respect to only $4 n_{\Lambda}$ variables. It turns out that the lowest energy trial wave functions satisfy $n_{\Lambda}\leq 3$ according to our variational calculations. The final results are shown in the phase diagram Fig.~\ref{fig:0}, and the details will be discussed in the following sections.

\section{Superfluids, Solids and Super-Solids}\label{sec3:phasediag}

\subsection{Static Structure Factors and Phase Diagram}
We use the static structure factor to characterize the long-range correlations of the trial wave functions. Among the eight generators, $\lambda^8$ and $\lambda^3, \lambda^4, \lambda^5$ are of importance since they generate the $U(2)$ symmetry group of the Hamiltonian (\ref{Ham}). Defining the vector $\boldsymbol{\lambda}^{\rm su2}=\lambda^4\ee_x+\lambda^5\ee_y+\lambda^3\ee_z$ (where $\ee_x,\ee_y,\ee_z$
are unit vectors)\textcolor{blue}{,} 
the static structure factors(SSFs) are given by 
\beq
    &&S_{\rm su2}(\boldsymbol{q})=\dfrac{1}{N}\sum_{i, \boldsymbol{\delta}} \expval{\pmb \lambda_i ^{\rm su2}\cdot \pmb \lambda_{i+\boldsymbol{\delta}}^{\rm su2}}
    \e^{-\ii\boldsymbol{q}\cdot\boldsymbol{\delta}}, \notag \\
    &&S_{8}(\boldsymbol{q})=\dfrac{1}{N}\sum_{i,\boldsymbol{\delta}} \expval{\lambda_i^8\lambda_{i+\boldsymbol{\delta}}^8}  \e^{-\ii\boldsymbol{q}\cdot\boldsymbol{\delta}}.
\eeq

\begin{figure}[t]
    \centering
    \includegraphics[width=1.0\linewidth]{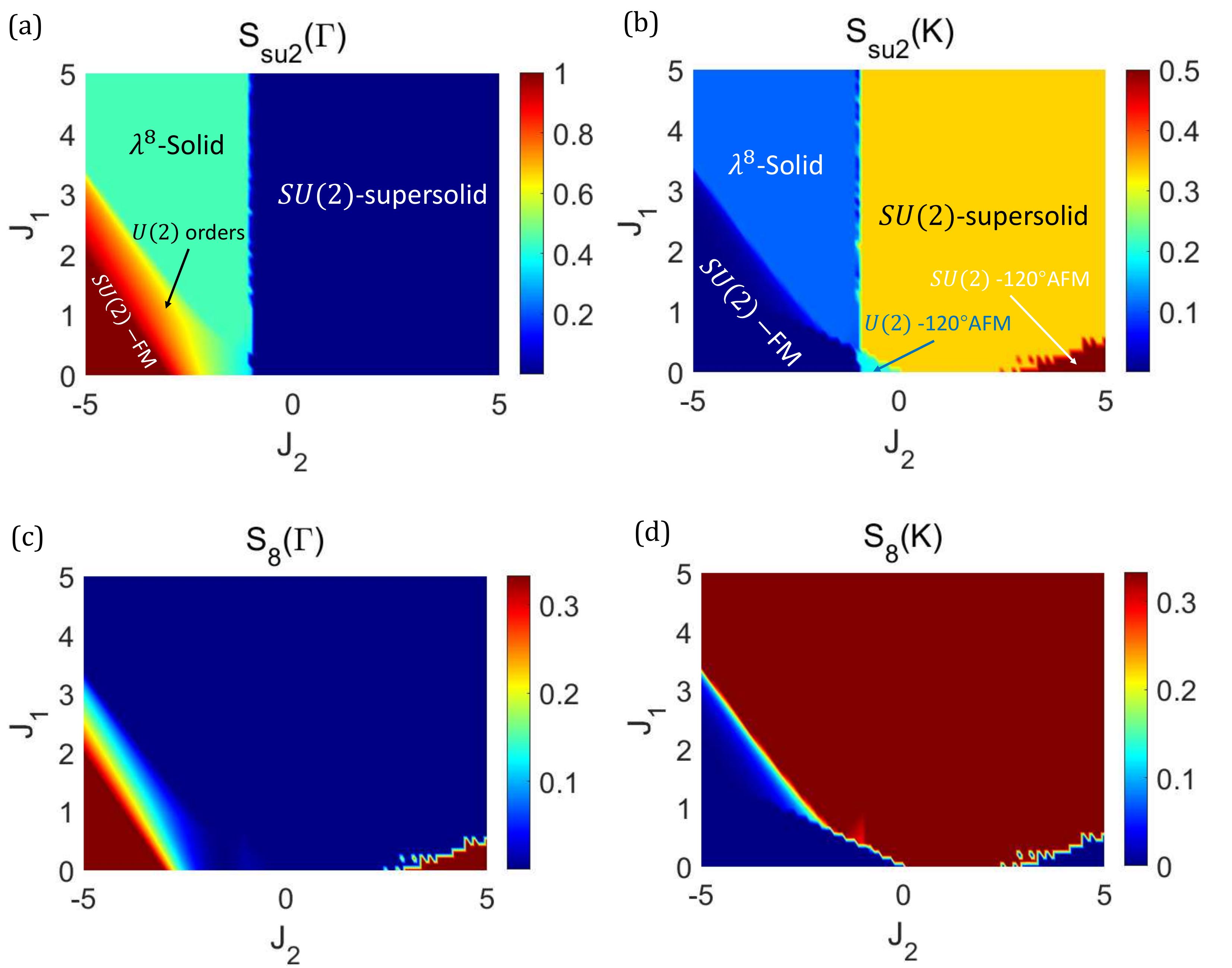}
    \caption{Distributions of the static structure factor (SSF) at the $\Gamma$ and $K$ points for different parameters. (a)$S_{\rm su2}(\Gamma)$. (b)$S_{\rm su2}(K)$. (c)$S_{8}(\Gamma)$. (d)$S_8(K)$.}
    \label{fig:2}
\end{figure}

\begin{figure}[t]
    \centering
    \includegraphics[width=1.\linewidth]{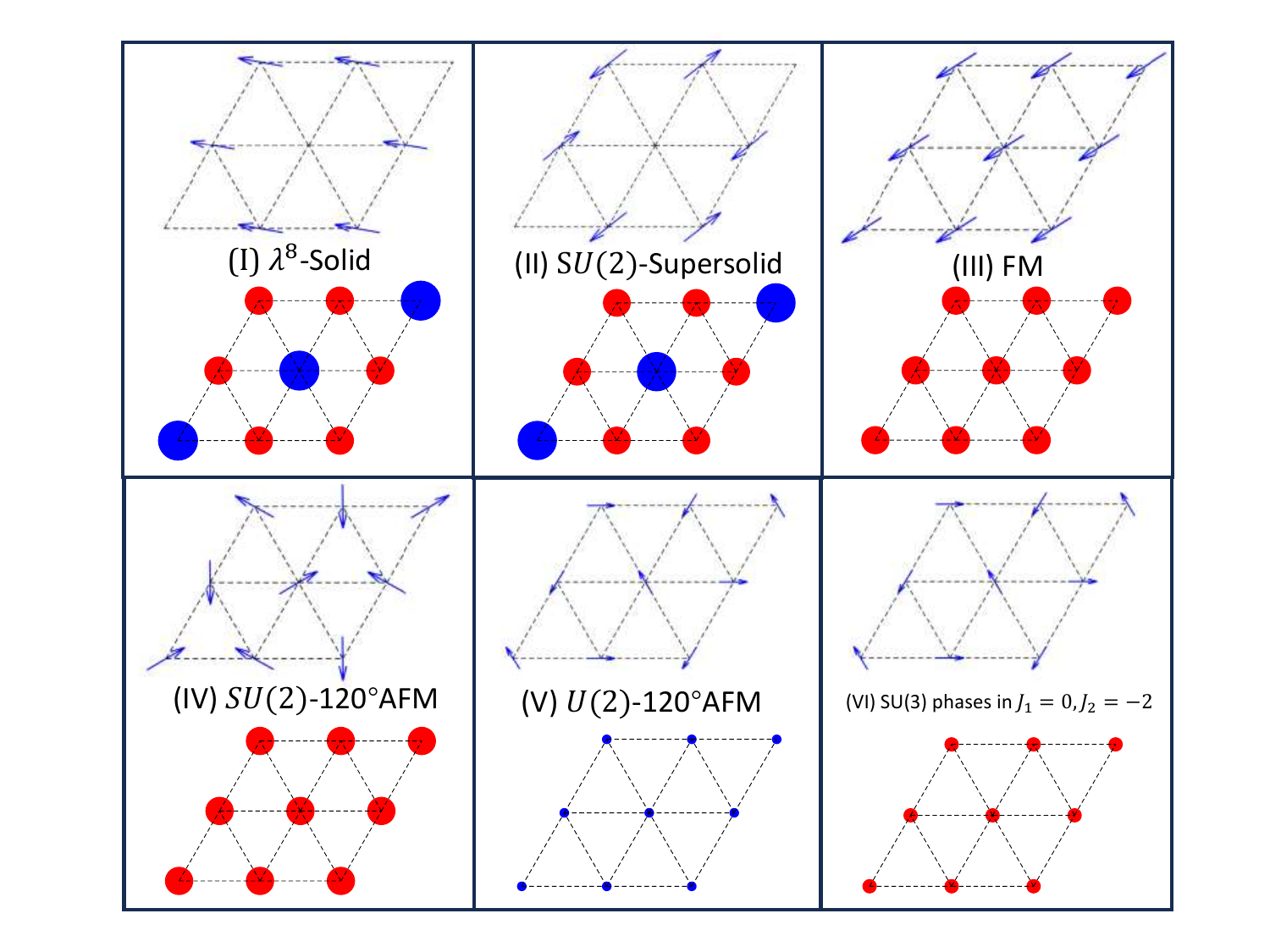}
    \caption{Distribution of $\expval{\boldsymbol{\lambda}^{\rm su2}}$ and $\lambda^8$ on the sublattices for the five order phases (I)-(VI) identified in Fig.\ref{fig:2}. The $\expval{\lambda^3}$ components of $\expval{\boldsymbol{\lambda}^{\rm su2}}$ are ommitted for clarity. The value of $\expval{\lambda^8}$ is indicated by 
    a solid dot on each site, with the radius representing the amplitude and red/blue color denoting the positive/negative sign. For the phase (VI), only the configuration at $J_1=0,J_2=-2$ has been exhibited.}
    \label{fig:3}
\end{figure}

We performed variational computation for the parameter region $-5\le J_2\le 5, 0\le J_1\le 5$ using (\ref{PsiN}) as the variational ansatz where the unit cell contains up to $n_\Lambda=6\times6$ lattice sites. Actually, the unit cell of all obtained ground states contains only $n_\Lambda\leq 3$ sites, hence we can use the SSF at ${\pmb q}=\Gamma, K$, as shown in in Fig.\ref{fig:2}-(a,b,c,d), to distinguish different phases. 
As shown in the phase diagram Fig.\ref{fig:0}, six phases are obtained labeled by I$\sim$VI. The pattern of various orders in these phases are illustrated in Fig.\ref{fig:3} by the expectation value of $\langle\boldsymbol{\lambda}^{\rm su2}\rangle$ and $\langle\lambda^8\rangle$ on each sublattice, where $\expval{\boldsymbol{\lambda}^{\rm su2}}$ is plotted as a three-dimensional vector, and $\expval{\lambda^8} $ is exhibited as a solid dot on each site with the radius representing the amplitude and the red/blue color denoting the positive/negative sign.

\subsection{Configuration of each phase}
\begin{figure*}
    \centering
    \includegraphics[width=1.0\linewidth]{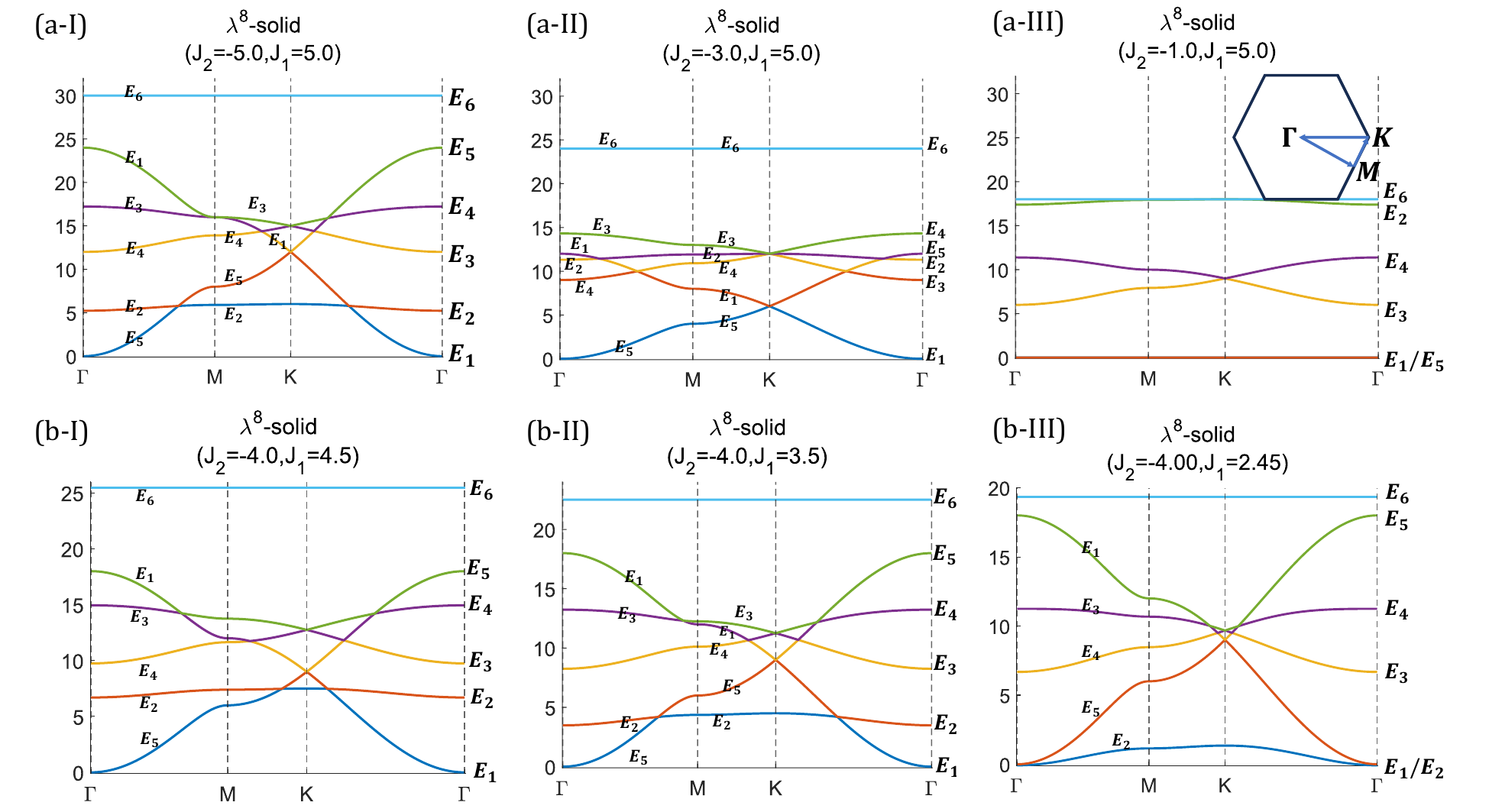}
    \caption{Spin wave spectrum for a typical $\lambda^8$-solid state. The magnon branch index $E_i$ is marked on the curves. (a)Spectra for $J_1=5.0$ and several values of $J_2$. (b)Spectra for $J_2=-4.0$ and several values of $J_1$. The inset of (a-III) shows the path taken in the first magnetic Brillouin zone for these calculations.}
    \label{fig:8}
\end{figure*}

Phase-I is characterized by the solid order of $\langle\lambda^8\rangle$ and exhibits a three-sublattice structure $(A,B,C)=(2\lambda,-\lambda,-\lambda)$. This phase is in analogy to the solid phase on the triangular lattice in hard-core boson systems\cite{Boninsegni_SuperSolidHCBosonTriangle_2005} and spin systems\cite{Blankschtein_frustratedTriangularSystem_1984}. Hence the SSF $S_8(\boldsymbol{q})$ exhibits a peak at the $K$ point. 
On the other hand, the magnitude of $\langle \boldsymbol{\lambda}^{\rm su2}\rangle$ is suppressed on $A$-sublattice where $\lambda^8$ has a larger value. On the other two sublattices $B$ and $C$, the value of $\langle\boldsymbol{\lambda}^{\rm su2}\rangle$ is `ferromagnetic' and forms a hexagonal superlattice. 
The non-zero net value of $\langle\boldsymbol{\lambda}^{\rm su2}\rangle$ in the unit cell induces an additional signal of the static structure factor at the $\Gamma$ point. Although the `ferromagnetic' order $\langle \boldsymbol{\lambda}^{\rm su2}\rangle$ breaks the $SU(2)$ symmetry, the $SU(3)$ linear spin-wave dispersion is quadratic (see Fig.\ref{fig:8}(a)) and cannot be considered as a `superfluid'. Hence we termed phase-I as a {\it $\lambda^8$-solid phase}.

It should be noticed that the $SU(2)$ group is generated by $\lambda^{3,4,5}$, where $\lambda^3$ is a magnetic dipole operator while $\lambda^{4,5}$ are quadrupole operators. Since the `ferromagnetic'  order $\langle\boldsymbol{\lambda}^{\rm su2}\rangle$ spontaneously breaks the $SU(2)$ symmetry, the resulting state generally contains coexisting magnetic dipole-order and quadrupole-order. The same situation also holds for all the other phases.

Phase-II is a very robust phase which contains a large volume in the phase diagram. It is also a solid phase of $\langle \lambda^8\rangle$ with a three-sublattice structure $(A,B,C)=(2\lambda,-\lambda,-\lambda)$. Furthermore, the value of $\langle\boldsymbol{\lambda}^{\rm su2}\rangle$ in the $SU(2)$ sector also vanishes on the A-sublattice. The difference is that the configuration of $\langle\boldsymbol{\lambda}^{\rm su2}\rangle$ form a Neel-type AFM order on the $B$- and $C$-sublattices . Hence the SSF $S_{\rm su2}$ exhibits a peak at the $K$ point. This structure is in analog to the `Y-phase' in the spin-1/2 systems\cite{Yamamoto_XXZ_2014,GaoYuan_SpinSuperSolidNBP_2022}. The collinear $\langle \boldsymbol{\lambda}^{\rm su2}\rangle $-magnetic order breaks the $SU(2)$ symmetry, and keep the lattice periodicity of the $\langle \boldsymbol{\lambda}^8\rangle$ order. Furthermore, the linear spin wave dispersion of the AFM phase is linear, indicating that it is indeed a `spin superfluid phase' which spontaneously breaks the $SU(2)$ symmetry to $U(1)$ (such that the total symmetry group reduces to $U(1)\times U(1)/Z_2$ with the first $U(1)$ generated by $\lambda^8$). Considering the coexistence of the solid order of $\langle \lambda^8\rangle$ and the superfluid order of $\langle \boldsymbol{\lambda}^{\rm su2}\rangle $, phase-II can be termed as a {\it $SU(2)$-supersolid phase} which has two Goldstone modes (see Fig.\ref{fig:9}(b)). It is different from the supersolid phase in literature which only breaks $U(1)$ spin rotation symmetry and contains only one Goldstone mode. 

Furthermore, the $SU(2)$ symmetry breaking only occurs at zero temperature. The finite temperature KT transition transition, which appears in $U(1)$-supersolid phase, is likely to be absent just as in the 2-dimensional classical Heisenberg model with $SO(3)$ symmetry\cite{Yao_HeisenbergModelPT_2025}. Nevertheless, a weak anisotropy in the $SU(2)$ sector can give rise to a KT phase at finite temperatures.


Phase-III is a uniform phase that does not break translation symmetry. In this phase, the distribution of $\expval{\lambda^8}$ and $\langle \boldsymbol{\lambda}^{\rm su2}\rangle$ are both `ferromagnetic' (FM), and only the $SU(2)$ symmetry is broken. The SSFs only exhibit peaks at the $\Gamma$ point. Hence phase-III is called a {\it $SU(2)$-FM phase}.

Phase-IV is characterized by the $120^{\circ}$-AFM order of $\langle \boldsymbol{\lambda}^{\rm su2}\rangle$ which exhibits a 3-sublattice structure. However, no solid order of $\expval{\lambda^8}$ is observed. 
It is in analog to the umbrella state found in spin-1/2 systems \cite{Yamamoto_XXZ_2014}. Therefore, this phase is not considered as a supersolid, but rather an {\it $SU(2)$-120$^\circ$AFM phase}. Since the $SU(2)$ symmetry is completely broken in the 120$^\circ$ AFM order, the spin wave dispersion contains three Goldstone modes (see Fig.\ref{fig:9}(a)).

\begin{table}[t]
    \centering
    \caption{Magnitude of $\lambda^{3,4,5,8}$ and $\lambda^{1,2,6,7}$ for phase (IV)$SU(2)$-AFM,and (V)$U(2)$-AFM. The parameters are $J_2=5,J_1=0$ for $SU(2)$-AFM and $J_2=-0.5,J_1=0$ for $U(2)$-AFM. Since the values are identical across all three sublattices for each phase, only the value for a single sublattice is listed}
      \begin{tabular}{c|c|c}
      \hline
            & $\sum\limits_{m=3,4,5,8}\expval{\lambda^m}^2$ & $\sum\limits_{m=1,2,6,7}\expval{\lambda^m}^2$ \\
      \hline
      $SU(2)$-AFM & 1.33  & 0 \\
      \hline
      $U(2)$-AFM & 0.4026 & 0.9307 \\
      \hline
      \end{tabular}%
    \label{tab:1}%
  \end{table}%

Phase-V also contains the $120^{\circ}$ AFM order for $\langle\boldsymbol{\lambda}^{\rm su2}\rangle$. But in this phase the value of $\langle \lambda^8\rangle$ is suppressed to almost zero on all sublattices. We further compare the quantities $\sum_{m=3,4,5,8} \expval{\lambda^m}^2$ and $\sum_{m=1,2,6,7}\expval{\lambda^m}^2$ in phase (IV) ($J_2=5,J_1=0$) and phase (V) ($J_2=-0.5, J_1=0$). We find that for each phase, these two values are identical on all three sublattices. However, as summarized in Tab.\ref{tab:1}, the value of $\langle\lambda^{1,2,6,7}\rangle$ is zero in phase (I)-(IV) (namely, all the orders only break the $SU(2)$ symmetry while preseve the $U(1)$ symmetry).
In contrast, in phase-V (and in region-VI discussed below), the value of $\sum_{m=1,2,6,7}\expval{\lambda^m}^2$ is significantly large, indicating that the order takes values in the whole eight generators of the $SU(3)$ group with complete breaking of the $U(2)$ symmetry. We therefore refer to phase (V) as {\it $U(2)$-AFM phase}.

The region-VI is sandwiched between the FM and the $\lambda^8$-solid and the value of $\sum_{m=1,2,6,7}\expval{\lambda^m}^2$ is also finite in this region. 
Actually, the region-VI contains more than one phase. For instance, a $\lambda^8$-Solid-like phase emerges in the parameter region close to $\lambda^8$-solid phase (for instance, $J_2=-5,J_1=3.3$). The difference between the two phases is that $\expval{\boldsymbol{\lambda}^{\rm su2}}$ is zero in the $A$-sublattice in the $\lambda^8$-solid phase but nonzero (but small) in the $\lambda^8$-solid-like phase. On the other hand, in the parameter region  between the FM (phase-III) and the $U(2)$-AFM (phase-V) (for instance, $J_2=-2.5,J_1=0$), 
a $U(2)$-FM-like phase emerges. The difference with the $SU(2)$-FM phase is that in the $U(2)$-FM-like phase, the weight of 
$\langle\lambda^8\rangle$ is suppressed and shifted to the remaining components $\expval{\lambda^{1,2,6,7}}$ (see Fig.\ref{fig:3}-(VI) for illustration). Generally, in region-VI the $U(2)$ generators are not sufficient to describe the orders any more. We will not go into much details in this region. 

\begin{figure}[t]
    \centering
    \includegraphics[width=1.03\linewidth]{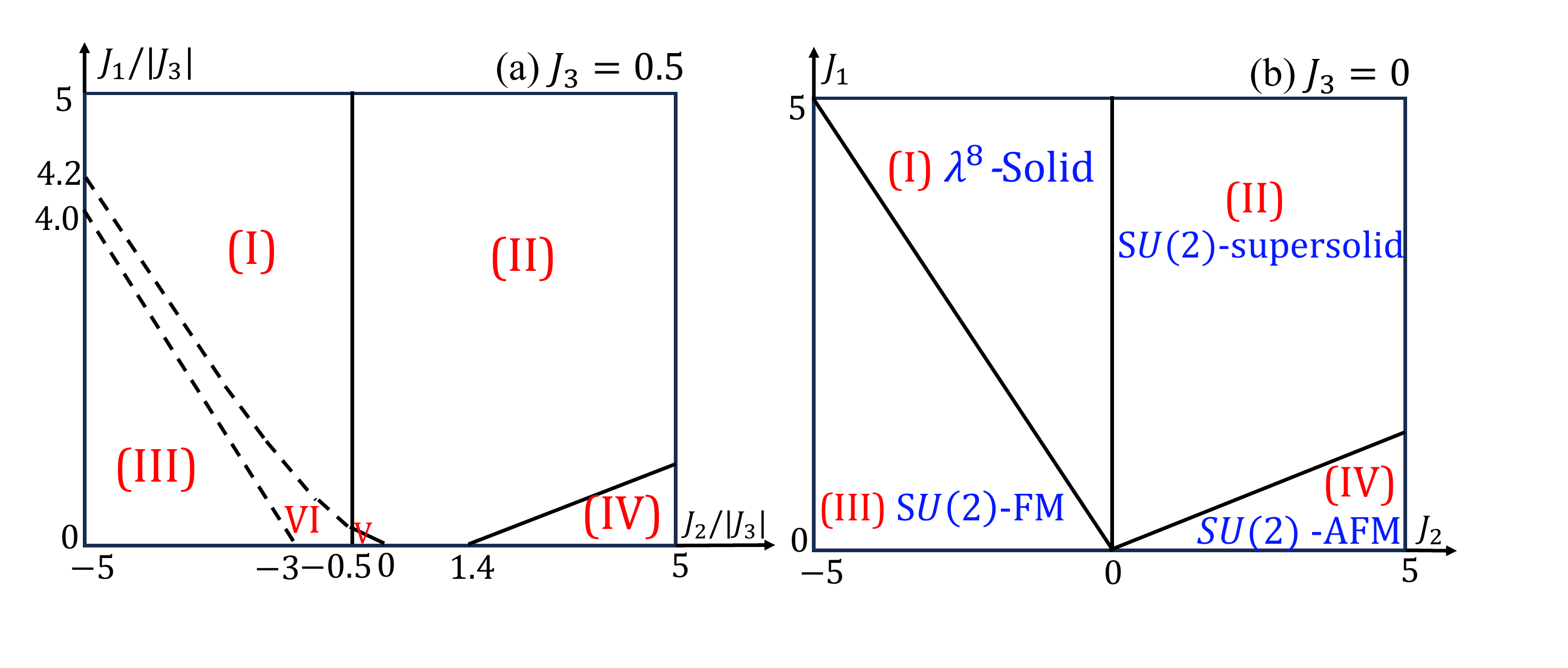}
    \caption{Variational phase diagram for the $U(2)$ model with different $SU(3)$ background coupling strengths. (a)$J_3=-0.5$. (b)$J_3=0$}
    \label{fig:4}
\end{figure}

Finally, the stability of the $U(2)$ ordered phases are determined by $|J_3|$, the amplitude of the $SU(3)$-symmetric interaction $H_{\rm su3}$. With the decrease of $|J_3|$, the extent of phase-V and region-VI reduces. Especially, when $J_3=0$, the phase boundary between the $\lambda^8$-solid and $SU(3)$ phases will become a straight line given by $|J_1|=|J_2|$, and both phase V and region-(VI) completely disappear (see Fig.\ref{fig:4}).

\subsection{phase transitions}

\begin{figure}[t]
    \centering
    \includegraphics[width=1.0\linewidth]{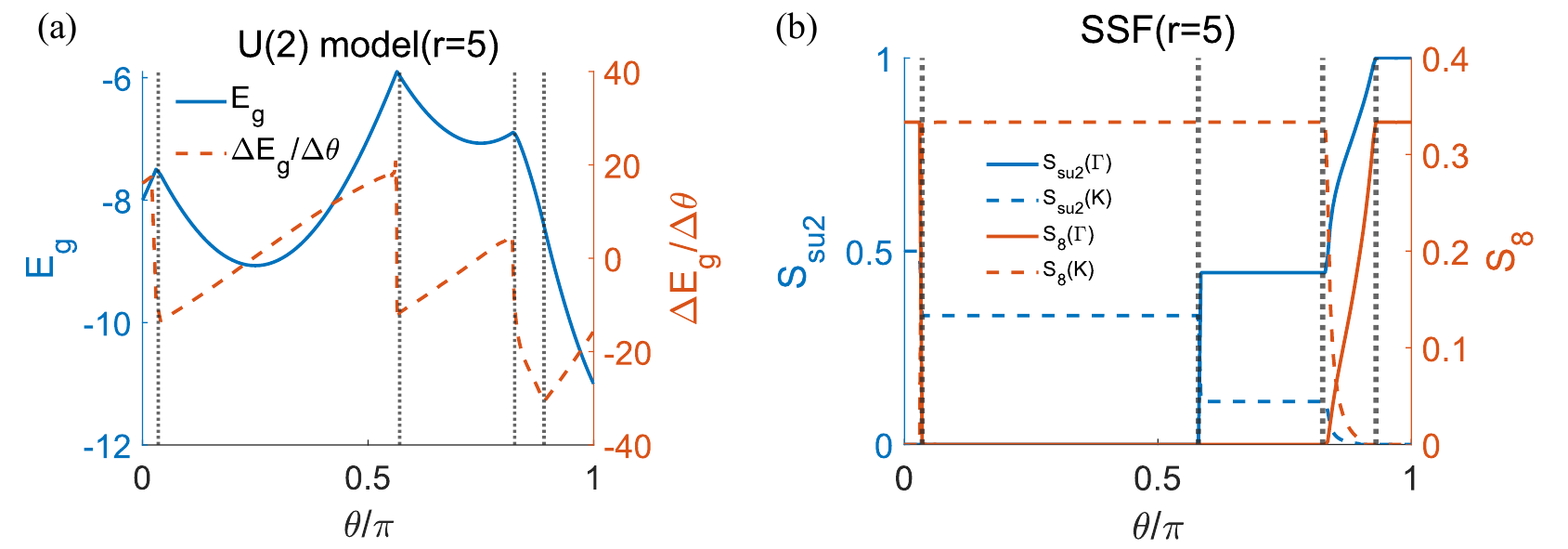}
    \caption{Variational computation results for different parameter $\theta=\arctan J_1/J_2$ with $r=\sqrt{J_1^2+J_2^2}=5$.(a)Energy of the variational ground state. The blue solid line represents the calculation results. The red scatter points are the first-order difference of energy. (b)Four SSF curves of the variational ground state. The solid line represents the value at the $\Gamma$ point while the dashed line represents the value of at the $K$ point. The blue curves represent the results of $S_{su2}(\boldsymbol{q})$, while the red curves represent the results of $S_8(\boldsymbol{q})$.}
    \label{fig:1}
\end{figure}

To illustrate the phase transitions, we consider a one-parameter curve $J_2=r\cos\theta, J_1=r\sin\theta$ with $r=5$ and $0\leq \theta\leq \pi$. This curve covers five different phases, along which two first-order phase transitions and two continuous phase transitions are observed (see Fig.\ref{fig:1}).

The SSFs and the first order derivative of the energy curve ${\partial E\over\partial \theta}$ exhibit discontinuous jumps at $\theta/\pi=0.035,0.58$, corresponding to the first-order phase transitions between phase-IV and phase-II, and between phase-II and phase-I, respectively. In contrast, the continuous change of SSFs and ${\partial E\over\partial \theta}$ at $\theta/\pi=0.825,0.93$, suggests a second-order transition between phase-I to region-VI, and between region-VI and phase-III, respectively.  
 
Within the interval $0.825<\theta/\pi<0.93$, all four SSF curves are non-zero and continuous, implying that the $SU(3)$-region-(VI) plays a role of a `transitional phase'. When setting $J_3=0$,  with the disappearance of the $SU(3)$-region-(VI), the direct transition between $\lambda^8$-Solid and FM becomes first-order.

The transition from phase-V to its neighboring phases (i.e. phase-II and region-VI), which are not covered by the above curve with $r=5$ and $0\leq \theta\leq \pi$, are both first order.

\section{$SU(3)$ spin-wave}\label{sec4:Su3spinwave}

To verify the stability of the classically ordered phases, we calculate the $SU(3)$ linear spin-wave\cite{Muniz_GeneralizedSpinWave_2014,Dong_SUNspin-wave_2018} in various regions of the phase diagram. Our results indicate that the edges for the spin-wave instabilities are in good consistency with the phase boundaries from the variational calculation, which indicates the validity of the semi-classical phase diagram.

\begin{figure*}
    \centering
    \includegraphics[width=1.0\linewidth]{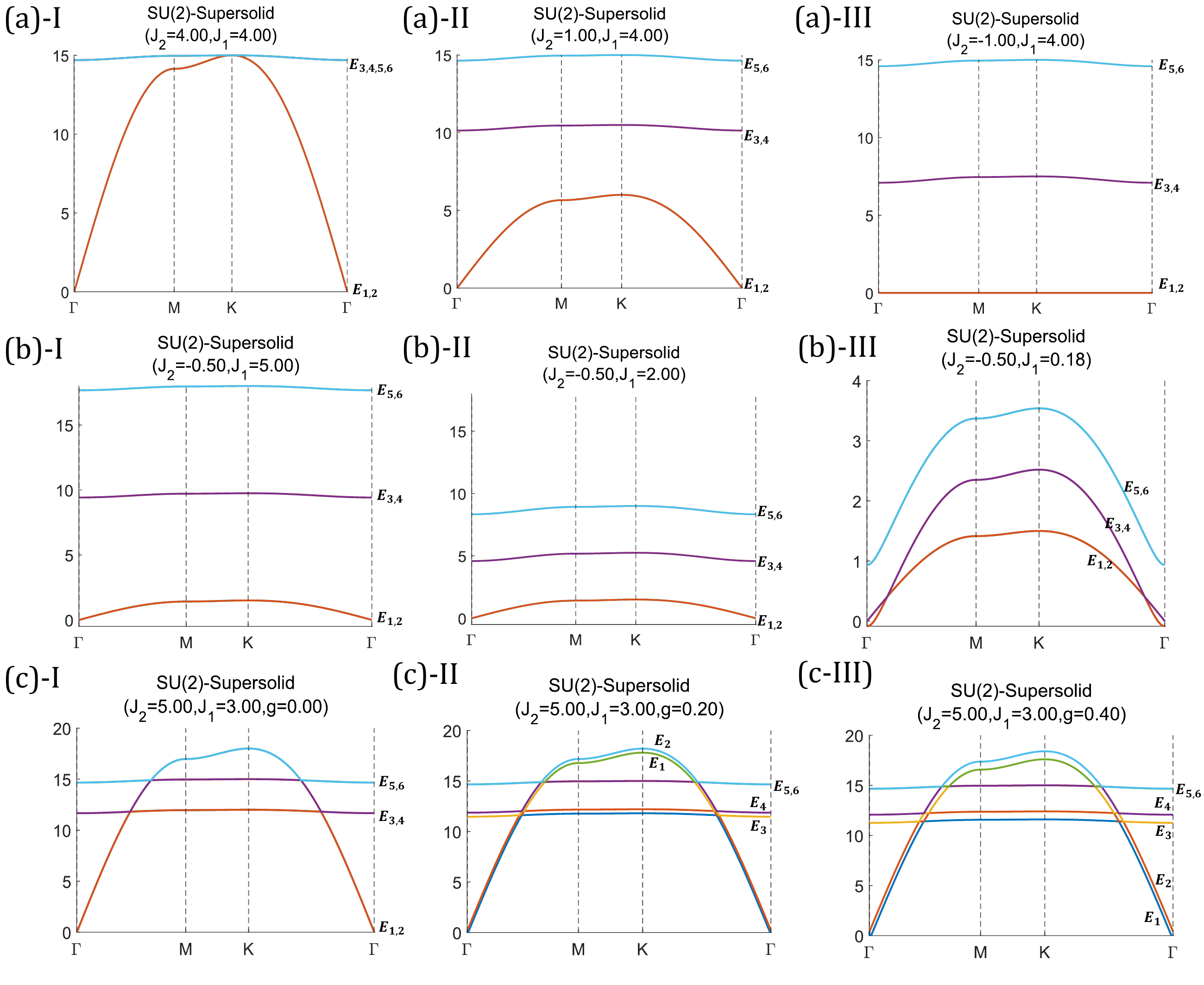}
    \caption{Spin-wave spectra for the $SU(2)$-supersolid state. The magnon branch index $E_i$ is marked on the curves. The calculation path is the same as Fig.\ref{fig:8}. Different series shows results for different parameter setting. (a)Spectra for $J_1=0.4$ and several values of $J_2$ with $SU(2)$-supersolid as reference state. (b)Spectra for $J_2=-0.5$ and several values of $J_1$ with $SU(2)$-supersolid as reference state. (c)Spectra for adding magnetic field coupling with $\lambda^3=S^z$. Here we denote $g$ as coupling strength, i.e. the model Hamiltonian has been modified as $H'=H+g\sum_i\lambda_i^3$.}
    \label{fig:9}
\end{figure*}

\subsection{$\lambda^8$-solid (phase-I)}
We consider a typical $\lambda^8$-Solid state with the configuration $\ket{\psi_A}=\ket{\psi_B}=\ket{1},\ket{\psi_C}=\ket{0}$. The results of the $SU(3)$ spin-wave calculation for this state at several parameter points are shown in Fig.\ref{fig:8}. The spectra $E_1,E_5$ are always gapless at the $\Gamma$ point and exhibit a quadratic dispersion relation, which is characteristic of ferromagnetism. With the decreasing of $J_2$, the bandwidth of $E_1, E_5$ becomes narrow, eventually turning to be completely flat and degenerate at zero energy when $J_2=-1.0$. If the magnitude of the antiferromagnetic $J_2$ is further reduced (i.e. $J_2>-1$), these two bands develop a negative quadratic form, indicating an instability and a phase transition from the ferromagnetic-like state to an antiferromagnetic state, as discussed previously.

On the other hand, if the strength of $J_2$ is fixed while parameter $J_1$ is reduced, the overall structure of the spectra remains largely unchanged. However, the band $E_2$ shifts downward and eventually closes its gap at the $\Gamma$ point near the transition. If the antiferromagnetic coupling $J_1$ is weakened further, $E_2$ develops negative excitation energies, signaling the instability of the FM order.

\subsection{$SU(2)$-supersolid (phase-II)}

Next, we calculate the spin-wave spectrum for the $SU(2)$-supersolid state (phase-II). Since the $SU(2)$ symmetry is spontaneously broken, the resulting $SU(2)$ order can be oriented along arbitrary direction. Without lose of generality, we assume that the $SU(2)$ order is carried by $\lambda^3$. (noticing that all of $\lambda^{3,4,5}$ reverse their sign under the action of $T$, the three generators are equivalent from the symmetry point of view). Then the classical ground state reads $\ket{\psi_A}=\ket{1},\ket{\psi_B}=\ket{-1},\ket{\psi_C}=\ket{0}$ respectively on the three sublattices.

Interestingly, the spin wave spectrum is doubly degenerate everywhere in the Brillouin zone (BZ). Hence only three distinct magnon dispersions are observed. Among the six magnon bands, the modes $E_1,E_2$ are gapless at the $\Gamma$ point and exhibit linear dispersion. This is a signature of the existence of spin superfluid order.  

Actually, the double degeneracy of the magnon spectrum is not accidental. In the ground state of the $SU(2)$ supersolid phase, the inversion symmetry $P$ and time reversal symmetry $T$ in (\ref{symT}) are broken, but the combined symmetry $PT$ with $(PT)^4=1$ is preserved (where the inversion center of $PT$ is located on the site of the $C$ sub-lattice). Furthermore, the $SU(2)$ symmetry generated by $\lambda^{3,4,5}$ is broken down to $U(1)=\{ e^{-i \lambda^3 \theta}; 0\leq\theta<2\pi \}$ generated by $\lambda^3=S_z$. Here we ignore the $U(1)$ symmetry generated by $\lambda^8$ since it does not contribute to the degeneracy. Since $PT$ anti-commutes with $\lambda_3$, the group formed by the remaining symmetries reads $G=U(1)\times Z_4^{PT}/Z_2$ with $Z_4^{PT}=\{E, PT, P_s, P_sPT\}, (PT)^2=P_s$, $Z_2=\{E,P_s\}$, where $P_s=e^{-i\lambda^3\pi}=e^{-i\lambda^4\pi}=e^{-i\lambda^5\pi}$ is the spin parity. Noticing that both the unitary elements $e^{-i\lambda^3\theta}$ and the anti-unitary element $PT$ do not change the lattice momentum, the spin-wave Hamiltonian at each $\pmb k$ point has the full symmetry $G=U(1)\times Z_4^{PT}/Z_2$. Furthermore, the magnon creation operators carry the 2-dimensional irreducible representations (irreps) in Tab.\ref{Irrep} (see Appendix \ref{app:irrep} for details), which yields the double degeneracy of the magnon bands throughout the BZ. This degeneracy can be lifted by perturbations that break the $G$ symmetry. For example, by applying a magnetic field coupling to $\lambda^3$, two of the three degenerate bands including the one containing two Goldstone modes, undergoes energy splitting (see Fig.\ref{fig:9}(c)). The splitting of the energy bands of the spin excitations are observable in experiments, and can be used to distinguish from the  $U(1)$-supersolid phase.

It is worthy mentioning that the two Goldstone modes created by $\lambda^4\pm i\lambda^5$ are not usual `magnons', because they carry quantum numbers $\pm 2$ of $S_z$, similar to the case in spin-1 XXZ model \cite{Sheng_QuadrupoleWaveSpinNematics_2025}. For this reason, the gapless Goldstone modes cannot be detected by neutron scattering experiment. However, if the $\pmb\lambda^{\rm su2}$ is oriented along different directions, then the Goldstone mode can contain both $S_z=\pm2$ and $S_z=0$ components, then can couple to neutron spin. On the other hand, by adopting the original $\lambda^{3,8}$ and redefining the $SU(2)$ symmetry of the model as generated by $\lambda^{1,2,3}$ (see Appendix~\ref{app:Lamd38}), then the Goldstone modes of the $SU(2)$-supersolid will generally be detectable in neutron scattering experiments.


\begin{table}[htbp]
    \centering
    \caption{A 2-dimensional irreducible representation of the group $G=U(1)\times Z_4^{PT}/Z_2$ carried by magnon creation operators.}\label{Irrep}
      \begin{tabular}{c|c|c|c}
      \hline
            & $e^{-i\lambda^3\theta}$ & $PT$ & bases \\
      \hline
        irrep-(1) & $\Bmat e^{-i\theta} &\\& e^{i\theta} \Emat$  &  $\Bmat  &-1 \\1 & \Emat K$ & $\begin{matrix}[(\lambda^{1}+i\lambda^{2}),  (\lambda^{6}-i\lambda^{7})],\\ [(\lambda^{6}+i\lambda^{7}),  -(\lambda^{1}-i\lambda^{2})].\end{matrix}$\\
      \hline
        irrep-(2) & $\Bmat e^{-2i\theta} &\\& e^{2i\theta} \Emat$  &  $\Bmat  &1 \\1 & \Emat K$ & $\begin{matrix} [(\lambda^{4}+i\lambda^{5}), -(\lambda^{4}-i\lambda^{5})]. \end{matrix}$\\
      \hline      
      \end{tabular}%
  \end{table}%

With $J_1$ fixed at $4.0$ (a strong antiferromagnetic coupling), we gradually decrease $J_2$ towards the ferromagnetic value $J_2=-1.0$ (see Fig.\ref{fig:9}(a)). The maximum excitation energy of magnons $E_1$ and $E_2$ decreases, eventually reaching zero accross the entire first magnetic BZ. A further increase in the ferromagnetic strength of $J_2$ (i.e., more negative value) results in globally negative excitation energies for $E_1,E_2$, indicating an instability and a phase transition. This behavior of the spectrum upon varying $J_2$ is similar to that observed in the $\lambda^8$-solid case, where bands $E_1,E_5$ become flat at $J_2=-1$ and exhibit trends analogous to bands $E_1$ and $E_2$ here. 

Next, we fix $J_2$ at a small ferromagnetic value $J_2=-0.5$ and gradually reduce $J_1$ (see Fig.\ref{fig:9}(b)). The shape of the dispersion curves for $E_1,E_2$ remains unchanged, but the energies of $E_3,E_4,E_5,E_6$ are shifted downward. At $J_1=0.18$, which corresponds to the transition point between the $SU(2)$-supersolid and $SU(3)$-$120^{\circ}$AFM phases, the gap of bands $E_3$ and $E_4$ closes. For values of $J_1$ below this critical point, parts of the  spectrum develop imaginary excitation energies, suggesting that the classically ordered $SU(2)$-supersolid state becomes unstable. 

Finally, when $J_2=J_3$, the bands $E_{3,4}$ will be degenerate with the bands $E_{5,6}$. If one further tunes $J_2=J_3=0$ such that the model is fully $SU(3)$ symmetric, then all six bands become completely degenerate in the whole Brillouin zone. This result is consistent with previous study of $SU(3)$ Heisenberg model in Ref.\cite{Bauer_SU3Heisenberg_2012}, where six-fold degeneracy of the excitation spectrum is observed in the reduced Brillouin zone.

\subsection{$SU(2)$-FM (phase-III)}
Then we calculate the spin-wave spectrum in the FM phase (phase-III). The FM order contains only one sublattice, hence two branches of magnons are expected from $SU(3)$ linear spin wave theory. 
Starting from the classical ground state $\ket{\psi_i}=\ket{1}$, one obtains the following two magnon bands with dispersion
\Beq
    &&E_1(\boldsymbol{k})= -2(J_3+J_2)\qty(6-\gamma_{\boldsymbol{k}}),\\
    &&E_2(\boldsymbol{k})= -2\qty[3(J_1+J_2) + J_3(6-\gamma_{\boldsymbol{k}})].
\Eeq
Here $\gamma_{\boldsymbol{k}}=\sum_{\boldsymbol{\delta}\in nn}\e^{-\ii\boldsymbol{k}\cdot\boldsymbol{\delta}}$ is the structural form factor. 
Parameterizing momentum as $\boldsymbol{k}=k_1\boldsymbol{b}_1+k_2\boldsymbol{b}_2$ (where $\boldsymbol{b}_1,\boldsymbol{b}_2$ are the reciprocal lattice vectors), then one obtains
\begin{align}
    \gamma_{\boldsymbol{k}}=2\qty(\cos 2\pi k_1+\cos 2\pi k_2+\cos 2\pi (k_1-k_2)).
\end{align}
Notice that the factor $(6-\gamma_k)\geq0$ is always valid. $(6-\gamma_k)$ equals 0 at the $\Gamma$ point and has a quadratic dispersion at small moments.
Furthermore, the magnon excitations must take non-negative energy, which requires that $E_1(\boldsymbol{k})\geq 0$, namely
\beq\label{FMcond1}
J_3+J_2<0.
\eeq
Violating of the above condition means the instability of the FM order. When (\ref{FMcond1}) is satisfied, then $E_1(\pmb k)$ is gapless at the $\Gamma$ point and have quadratic magnon dispersion, which is the signature feature of the FM order.

On the other hand, the other band $E_2(\pmb k)$ is generally gapped, hence $E_2(\boldsymbol{k}) > 0$ should be valid in the whole Brillouin zone. It requires that $3(J_1+J_2) + J_3(6-\gamma_{\boldsymbol{k}})<0$, namely $J_1+J_2<{J_3\over3}(\gamma_k-6)$ for any $\pmb k$. Since $(\gamma_k-6)$ takes its minimum value $-3$ at the $K$ point, one should have
\beq\label{FMcond2}
J_1+J_2<-3J_3.
\eeq
At the parameter line $J_1+J_2=-3J_3$, the $E_2(\pmb k)$ band becomes gapless at the $K$ point, namely, $E_2(K)=0$. This indicates that the magnons will undergo BEC. The condensation of magnons with momentum $K$ indicates a phase transition to another ordered phase with three sublattices. 

Especially, if $J_3=0$, the phase transition line becomes $J_1 = -J_2$. These results are consistent with our earlier variational calculations.

\subsection{$SU(2)$-120$^\circ$AFM (phase-IV)}

\begin{figure*}
    \centering
    \includegraphics[width=1.0\linewidth]{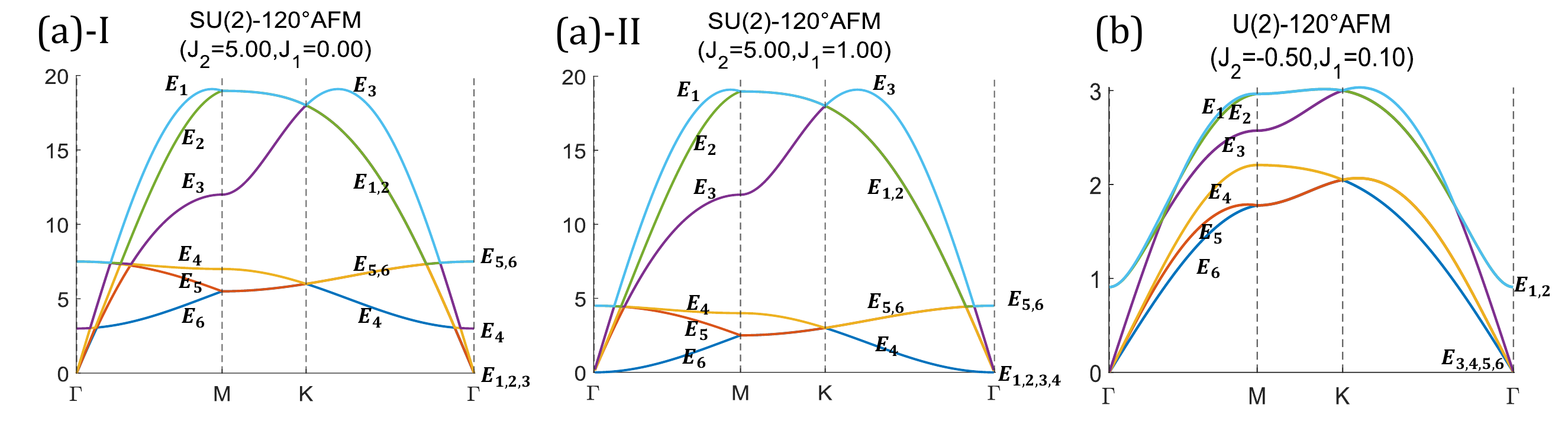}
    \caption{Spin-wave spectra for the $SU(2)$-$120^{\circ}$AFM state, $U(2)$-$120^{\circ}$AFM state. The magnon branch index $E_i$ is marked on the curves. The calculation path is the same as Fig.\ref{fig:8}. Different series shows results for different setting. (a)Spectra for $J_2=5.0$ and two values of $J_1$ with $SU(2)$-$120^{\circ}$AFM as reference state. (b)Spectra for $J_2=-0.5,J_1=0.1$ with $U(2)$-$120^{\circ}$AFM as reference state. }
    \label{fig:10}
\end{figure*}

We performed a variational calculation for the parameters $J_2=5.0,J_1=0.0$ to obtain the components of $\ket{\psi_A},\ket{\psi_B},\ket{\psi_C}$ in the $S^z$ basis. This state servered as the reference state for the spin-wave calculation.

The six magnon spectra can be classified into two sets. The first set consists of three gapless modes $(E_1,E_2,E_3)$ whose gaps close at the $\Gamma$ point. They exhibit a linear dispersion relation nearby, which is a typical signature of antiferromagnetism. Those three branches of linearly dispersive gapless magnons correspond to the three Goldstone modes arising from the complete breaking of the $SU(2)$ symmetry. 

The other three magnons are gapped across most of the $SU(2)$-$120^{\circ}$AFM phase region. Their excitation energies shift overall with changes in $J_1$ and $J_2$. As shown in Fig.\ref{fig:10}(a), the energies of these modes decrease as the antiferromagnetic coupling $J_1$ strengthens, eventually becoming gapless at the phase boundary, indicating a phase transition.

\subsection{$U(2)$-AFM (phase-V)}

Finally, we  calculate the spin-wave spectrum for the $U(2)$-AFM state (phase-V) at $J_2=-0.5,J_1=0.1$ (see Fig.\ref{fig:9}(d)), with the reference configuration chosen from the results of variational calculation. Since the nonzero values of $\langle\lambda^{1,2,6,7}\rangle$ completely break the $U(2)$ symmetry, the spin-wave dispersion contains four branches of gapless Goldstone modes and the rest two are gapped.
If we change the parameters to another point within the $SU(3)$-AFM region but keep the reference configuration unchanged, imaginary spectra will appear near the $\Gamma$ point, which implies quantum fluctuations are strong in this phase.

\section{Conclusion and Discussion} \label{sec5:sum}
In summary, we studied a $U(2)=SU(2)\times U(1)/Z_2$ symmetric spin-1 model on the triangular lattice. By employing a variational calculation within a classical $\mathbb CP^2$ ansatz, we determined a rich phase diagram that contains various long-range ordered phases. The most significant finding of this work is the identification of a robust $SU(2)$-supersolid phase. In this phase, the system spontaneously breaks both the spin $SU(2)$ symmetry---via a non-collinear magnetic order of $\boldsymbol{\lambda}^{\rm su2}$---and the lattice translation symmetry---via a solid order parameter $\lambda^8$ with a 3-sublattice $(2,-1,-1)$ pattern. This represents a direct generalization of the supersolid phenomenon to a higher spin context with a more complex broken symmetry group. The stability of this classically ordered state and its characteristic excitation spectrum, including intriguing degeneracies, were further validated by $SU(3)$ linear spin-wave theory.

Beyond the supersolid, our study unveils a few other phases. These include conventional FM or $120^{\circ}$-AFM phases within the $SU(2)$ sector, where translational symmetry remains intact or is broken by the magnetic order itself. Notably, in regions where $J_1,J_2$ are weak, the non-zero value of $\sum_{m=1,2,6,7}\expval{\lambda^m}^2$ implies a more exotic ordered state which completely breaks the $U(2)$ symmetry. The nature of these phases warrants further investigation.

Our work establishes the designed $U(2)$ model as a fertile ground for exploring complex magnetic orders and multiple symmetry breaking in spin-1 systems.  The natural next step involves confronting these classical and spin-wave results with fully quantum-mechanical numerical simulations to assess the impact of quantum fluctuations. Furthermore, it would be intriguing to explore the potential realization of these phases in realistic magnetic compounds or quantum simulators featuring spin-1 moments on triangular lattices.

\section*{ACKNOWLEDGEMENTS}
We thank Qiang Luo, Han-Jin Shen, Rong Yu and Zhi-yuan Xie for valuable discussions, and thank Yi Zhou and T.K. Ng for previous collaborations in related topics. This work was supported by National Basic Research and Development plan of China (Grants No.2023YFA1406500, 2022YFA1405300) and NSFC (Grants No.12374166,12134020). Computational resources have been provided by the Physical Laboratory of High Performance Computing at Renmin University of China.

\appendix
\section{Details of Spin wave calculations}
\subsection{Spin wave of FM state}
Let $\ket{1},\ket{0},\ket{-1}$ denote the eigenstates in the $\hat S^z$ basis, in which Gell-Mann matrices take their standard form. Without loss of generality, the FM state $\ket{FM}$ can be chosen as $\ket{FM}=\prod_i\ket{1}_i$. Introducing Schwinger Boson via the mapping $\hat b_{1}\ket{\rm vac}\cong\ket{1},\hat b_{0}\ket{\rm vac}\cong\ket{0},\hat b_{-1}\ket{\rm vac}\cong \ket{-1}$\cite{Arovas_SchwingerBoson_1988}, the spin-wave theory treatment assumes the condensation of the $\hat b_{1}$ boson, while $\hat b_{0},\hat b_{-1}$ become the Holstein-Primakoff(HP) bosons. The Gell-Mann matrices can then be rewritten in terms of Schwinger bosons. The operator $\hat b_{1}$ is approximated as\begin{align}
    \hat b_{1}^{\dagger}\approx \hat b_{1}\approx \sqrt{1-\hat b_{0}^{\dagger}\hat b_0-\hat b_{-1}^{\dagger}\hat b_{-1}}
\end{align}
reflecting the constraint on the total boson number per site. Within the linear spin-wave approximation, the single operators $\hat b_{1},\hat b_{1}^{\dagger}$ are replaced by $1$, while their product is substituded using the number constraint:\begin{align}
    b_{1}^{\dagger}\hat b_{1}=1-\hat b_0^{\dagger}\hat b_0-\hat b_{-1}^{\dagger}\hat b_{-1}
\end{align}
Expanding the Hamiltonian and retaining terms up to quadratic order in the HP boson operators yields the general form\begin{align}
    \hat H=&\sum_{\expval{ij}}\qty(T_{\alpha\beta}\hat b_{i\alpha}^{\dagger}\hat b_{j\beta}+\Delta_{\alpha\beta}\hat b_{i\alpha}^{\dagger}\hat b_{j\beta}^{\dagger}+h.c.)\\
    &+12\sum_{i}V_{\alpha\beta}\hat b_{i\alpha}^{\dagger}\hat b_{i\beta}+E_0
\end{align}
where the indices $\alpha,\beta$ run over the two flavors of HP bosons ($0$ and $\down$), and the matrices are defined by\begin{align}
    \begin{cases}
        V_{\alpha\beta}=\sum_{m=1}^8J\level{m}\lambda_{00}^m\qty(\lambda_{\alpha\beta}^m-\lambda_{00}^m\delta_{\alpha\beta})\\
        T_{\alpha\beta}=\sum_{m=1}^8J\level{m}\lambda_{\alpha 0}^m\lambda_{0\beta}^m\\
        \Delta_{\alpha\beta}=\sum_{m=1}^8J\level{m}\lambda_{\alpha 0}^m\lambda_{\beta 0}^m
    \end{cases}
\end{align}
For our model Hamiltonian, the coupling strengths $J\level{m}$ are assigned as\begin{align}
    J\level{m}=\begin{cases}
        J_2+J_3\qquad m=3,4,5\\
        J_3\qquad m=1,2,6,7\\
        J_1+J_3\qquad m=8
    \end{cases}
\end{align}
Performing a Fourier transformation $\hat b_{\boldsymbol{k}\alpha}=\dfrac{1}{\sqrt{N}}\sum_i\hat b_{i\alpha}\e^{-\ii\boldsymbol{k}\cdot\boldsymbol{R}_i}$ and diagonalizing the Hamiltonian matrix at each momentum point yields the magnon spectrum.

\subsection{Spin wave for 3-sublattice state}

We now introduce spin-wave theory for generic 3-sublattice states. In this case, the spin state on the three sublattices (A,B,C) may differ and are generally not the standard $\ket{1}$ state. Therefore, a separate local basis rotation must be performed on each sublattice such that its respective ground state spin configuration aligns with the first basis vector of the local Hilbert space. In these rotated basis, the Gell-Mann matrices assume representations different from the standard one. We denote these new representations for sublattice $S$ as $\lambda^{mS}(m=1,\cdots,8; S=A,B,C)$. We then introduce three flavors of Schwinger bosons, $\hat b_{iA\alpha},\hat b_{iB\alpha},\hat b_{iC\alpha}$ corresponding to each sublattice. The model Hamiltonian of model can be decomposed into three parts based on the three types of nearest-neighbor bonds:\begin{align}
\hat H=\hat H_{AB}+\hat H_{BC}+\hat H_{CA}
\end{align}
Each part has an identical strcture. Using $\hat H_{AB}$ as an example, we have\begin{align}
    \hat H_{AB}=&\sum_{\expval{ij}_{AB}}\qty(T_{\alpha\beta}^{AB}\hat b_{iA\alpha}^{\dagger}\hat b_{jB\beta}+h.c.)\\
    +&\sum_{\expval{ij}_{AB}}\qty(\Delta_{\alpha\beta}^{AB}\hat b_{iA\alpha}^{\dagger}\hat b_{jB\beta}^{\dagger}+h.c.)\\
    +&z_{AB}\sum_{i\in A}V_{\alpha\beta}^{AB}\hat b_{iA\alpha}^{\dagger}\hat b_{iA\beta}\\
    +&z_{AB}\sum_{i\in B}V_{\alpha\beta}^{BA}\hat b_{iB\alpha}^{\dagger}\hat b_{iB\beta}+E_0
\end{align}
Here, $z_{AB}=3$ is the coordination number for an AB-bond, and matrices $\boldsymbol{V}^{AB},\boldsymbol{\Delta}^{AB},\boldsymbol{T}^{AB}$ are defined as\begin{align}
    \begin{cases}
        V_{\alpha\beta}^{AB}=\sum_{m=1}^8J\level{m}\qty(\lambda_{\alpha\beta}^{mA}-\lambda_{00}^{mA}\delta_{\alpha\beta})\lambda_{00}^{mB}\\
        T_{\alpha\beta}^{AB}=\sum_{m=1}^8J\level{m}\lambda_{\alpha 0}^{mA}\lambda_{0\beta}^{mB}\\
        \Delta_{\alpha\beta}^{AB}=\sum_{m=1}^8J\level{m}\lambda_{\alpha 0}^{mA}\lambda_{\beta 0}^{mB}
    \end{cases}
\end{align}
After performing the Fourier transformation $\hat b_{\boldsymbol{k}S\alpha}=\sqrt{\dfrac{3}{N}}\sum_{i}\hat b_{iS\alpha}\e^{-\ii\boldsymbol{k}\cdot\boldsymbol{R}_{i}}$ (where $S=A,B,C$), the Hamiltonian can be expressed as a $12\times 12$ matrix $\mathcal{H}(\boldsymbol{k})$ in the basis of these bosonic operators. To diagonlize this matrix, one must solve the Bogoliubov-de Gennes eigenvalue problem $\Sigma\mathcal{H}(\boldsymbol{k})\boldsymbol{v}_i=E_i(\boldsymbol{k})\boldsymbol{v}_i$, where the metric $\Sigma$ is given by\begin{align}
    \Sigma=\mqty(\boldsymbol{I}_6&\\&-\boldsymbol{I}_6)
\end{align}
The six eigenvalues $E_i(\boldsymbol{k})$ found at each momentum point correspond to the spin-wave excitation energies. Their associated eigenvectors $\boldsymbol{v}_i$ must satisfy the positive normalization condition $\boldsymbol{v}_i^{\dagger}\Sigma\boldsymbol{v}_i=1$. These six modes correspond to two magnon branches originating from each of the three sublattices.

\section{Irreps of the group $G=U(1)\times Z_4^{PT}/Z_2$} \label{app:irrep}

The group $G=U(1)\times Z_4^{PT}/Z_2$ has one 1-dimensional irrep and infinite 2-dimensional irreps, as shown in Tab.\ref{Irreps}. Each irrep is labeled by an integer $m\in \mathbb Z$, namely the (abstract) value of the $U(1)$ `charge'. 

\begin{table}[htbp]
    \centering
    \caption{Irreducible representation of the group $G=U(1)\times Z_4^{PT}/Z_2$ carried by magnon creation operators. Notice that $(PT)^2=(-1)^m$ is the spin parity (quantum number of $P_s=e^{-i\lambda^3\pi}$) with $m$ the eigenvalue of $S_z=\lambda^3$.}\label{Irreps}
      \begin{tabular}{c|c|c|c}
      \hline
            & $e^{-i\lambda^3\theta}$ & $PT$ & dimensionality \\
      \hline
        irrep-(0)  & 1  &  $1 K$ &1  \\
       \hline
        irrep-(1)  & $\Bmat e^{-i\theta} &\\& e^{i\theta} \Emat$  &  $\Bmat  &-1 \\1 & \Emat K$ &2\\
       \hline	 
	\vdots & \vdots & \vdots & \vdots \\
      \hline
	 irrep-($m$)  & $\Bmat e^{-im\theta} &\\& e^{im\theta} \Emat$  &  $\Bmat  &(-1)^m \\1 & \Emat K$  &2 \\
      \hline      
	\vdots & \vdots & \vdots& \vdots \\
      \hline
      \end{tabular}%
  \end{table}%

In the following, we illustrate that the spin-wave operators in the $SU(2)$ super-solid phase carry a 2-dimensional irrep labeled by $m=1$. 

The $SU(2)$ super-solid phase is a three-sub-lattice order having the ground state $\ket{\psi_A}=\ket{1},\ket{\psi_B}=\ket{-1},\ket{\psi_C}=\ket{0}$.  According to $SU(3)$ spin wave theory, each sublattice contributes two magnon bands. On sublattice-A ($\ket{1}$), the magnon creation operators are given by 
\[
{1\over2}(\lambda^1-i\lambda^2) = \Bmat 0&0&0 \\1&0&0\\0&0&0\Emat,\  
{1\over2}(\lambda^4-i\lambda^5) = \Bmat 0&0&0 \\0&0&0\\1&0&0\Emat.  
\] 
Similarly, on sublattice-B ($\ket{-1}$), the magnon creation operators are
\[
{1\over2}(\lambda^6+i\lambda^7) = \Bmat 0&0&0 \\0&0&1\\0&0&0\Emat,\ 
{1\over2}(\lambda^4+i\lambda^5) = \Bmat 0&0&1 \\0&0&0\\0&0&0\Emat.
\] 
and on sublattice-C ($\ket{0}$),
\[
{1\over2}(\lambda^1+i\lambda^2) = \Bmat 0&1&0 \\0&0&0\\0&0&0\Emat,\  
{1\over2}(\lambda^6-i\lambda^7) = \Bmat 0&0&0 \\0&0&0\\0&1&0\Emat.
\] 


These magnon operators carry nontrivial quantum numbers of the $U(1)=\{e^{-i\lambda^3\theta}\}$ group. For instance, 
\beq
[\lambda^3, (\lambda^1\pm i\lambda^2)]&=&\lambda^3 (\lambda^1\pm i\lambda^2) - (\lambda^1\pm i\lambda^2) \lambda^3 \notag\\
&=& \pm (\lambda^1\pm i\lambda^2),
\eeq
so ${1\over2}(\lambda^1+i\lambda^2)$ and ${1\over2}(\lambda^1-i\lambda^2)$ carry $U(1)$ `charge' $+1$ and $-1$ respectively, namely
\[
e^{-i\lambda^3\theta} (\lambda^1\pm i\lambda^2) e^{i\lambda^3\theta} = e^{\mp i\theta}(\lambda^1\pm i\lambda^2).
\]

Similarly,  $(\lambda^6\pm i\lambda^7)$ carry $U(1)$ `charge' $\pm 1$. However, because $[\lambda_3, (\lambda^4\pm i\lambda^5)] =\pm2 (\lambda^4\pm i\lambda^5)$,  so $(\lambda^4\pm i\lambda^5)$ carry `charge' $\pm 2$, respectively.  

On the other hand, under the action of $PT$, the magnon creation operators form pairs. Namely,
\Beq
&&(PT) (\lambda^1\pm i\lambda^2) (PT)^{-1} = +(\lambda^6\mp i\lambda^7),\\
&&(PT) (\lambda^6\pm i\lambda^7) (PT)^{-1} = -(\lambda^1\mp i\lambda^2),\\
&&(PT) (\lambda^4+i\lambda^5) (PT)^{-1} = -(\lambda^4-i\lambda^5),\\
&&(PT) (\lambda^4-i\lambda^5) (PT)^{-1} = -(\lambda^4+i\lambda^5).
\Eeq

Therefore, the pair of magnon operators $[(\lambda^1 + i\lambda^2),  (\lambda^6- i\lambda^7)]$ carries the 2-dimensional irrep $(m=1)$ of the group $G=U(1)\times Z_4^{PT}/Z_2$.  Similarly, the bases $[(\lambda^6 + i\lambda^7),  -(\lambda^1- i\lambda^2)]$ also carry the same irrep of $G$. But $[(\lambda^4+i\lambda^5), -(\lambda^4-i\lambda^5)]$ carried the irrep $(m=2)$.

\section{Alternative definition of the $U(2)$ symmetry group}
\label{app:Lamd38}
In this work, we have modified the definition of $\lambda^3,\lambda^8$, to make sure that $\lambda^3$ fully represents the magnetic moment in the general convention. Under this convention, $\ket{\pm 1}$, two bases of the sub-Hilbert space for single spin-1 site which can be connected with time-reversal operation, can take the role of $SU(2)$ generators, acting on the subspace $\text{Span}\qty{\ket{1}, \ket{-1}}$. This modification of Gell-Mann matrices may be more realistic to real magnetic systems, and become a key factor in the emergence of spin-2 excitations. 

On the other hand, one can adopt the origin convention of $\lambda^{3,8}$, with $\lambda^3={1\over2}S_z+{3\over2}S_z^2-I, \lambda^8 = ({3\over2}S_z -{3\over2}S_z^2 +I)/\sqrt3$ namely,
\Beq
\lambda^3=\Bmat1&0&0\\0&-1&0\\0&0&0\Emat, \lambda^8={1\over\sqrt3}\Bmat1&0&0\\0&1&0\\0&0&-2\Emat,
\Eeq
and redefine the $SU(2)$ symmetry group to be generated by $\lambda^{1,2,3}$ which act on the subspace $\text{Span}\qty{\ket{1}, \ket{0}}$. Then the model Hamiltonian is modified as
\Beq
H = J_3\sum_{\expval{ij}}\sum_{\alpha=1}^8\lambda^\alpha_{i}\lambda^\alpha_{j}+ J_2\sum_{\expval{ij}}\sum_{m=1}^3\lambda_{i}^m\lambda_{j}^m + J_1\sum_{\expval{ij}}\lambda^8_{i}\lambda^8_{j},
\Eeq
and the excitations including the Goldstone modes can carry $S_z$ quantum numbers $\pm1$, and then can be detected in neutron scattering experiments. 

\bibliographystyle{iopart-num.bst}
\bibliography{Ref.bib}

\end{document}